\documentclass[aps,preprintnumbers,twocolumn]{revtex4}
\usepackage{hyperref}
\usepackage{amsmath}
\usepackage{amsfonts}
\usepackage{amssymb}
\usepackage{graphicx}
\usepackage{empheq}
\usepackage{lipsum}

\hypersetup{colorlinks=true, urlcolor=blue, linkcolor=blue, citecolor=blue,plainpages=false}

\newcommand{\partialdiff}[2]{\frac{\partial{#1}}{\partial{#2}}}
\newcommand{\txtpow}[1]{{\mbox{\scriptsize{#1}}}}

\newcommand{\boldeta}{\boldsymbol{\eta}}
\renewcommand{\eqref}[1]{{Eq.~(\ref{#1}})}

\newcommand{\revision}[1]{{{#1}}}

\DeclareSymbolFont{bbold}{U}{bbold}{m}{n}
\DeclareSymbolFontAlphabet{\mathbbold}{bbold}

\begin{document}

\title{Universal bounds for imaging in scattering media}

\author{Niall Byrnes$^1$}
\author{Matthew R. Foreman$^{1}$}
\email[]{matthew.foreman@imperial.ac.uk}
\affiliation{Blackett Laboratory, Department of Physics, Imperial College London, Prince Consort Road, London, SW7 2AZ, UK}

\date{\today}

\begin{abstract}
In this work we establish universal ensemble independent bounds on the mean and variance of the mutual information and channel capacity for imaging through a complex medium. Both upper and lower bounds are derived and are solely dependent on the mean transmittance of the medium and the number of degrees of freedom $N$. In the asymptotic limit of large $N$, upper bounds on the channel capacity are shown to be well approximated by that of a bimodal channel with independent identically Bernoulli distributed transmission eigenvalues. Reflection based imaging modalities are also considered and permitted regions in the transmission-reflection information plane defined. Numerical examples drawn from the circular and DMPK random matrix ensembles are used to illustrate the validity of the derived bounds. Finally, although the mutual information and channel capacity are shown to be non-linear statistics of the transmission eigenvalues, the existence of central limit theorems is demonstrated and discussed.
\end{abstract}

\maketitle 
	
\section{Introduction}\label{sec:intro}

The need to image through a complex scattering medium occurs frequently in, for example,  biomedical optics, aerial reconnaissance, remote sensing and astronomy \cite{Tuchin2015,Zege1991,Hansen1974}. Such efforts are, however, frequently impeded since upon transmission through complex media the structure of the incident image field is strongly modified resulting in a randomly varying output speckle pattern bearing little or no resemblance to the original image. To overcome this problem numerous techniques have been developed in recent years. Measurement of the transmission matrix of a scattering medium, for example, allows retrieval of the original image by application of the associated inverse operation \cite{Popoff2010a,Popoff2010,Popoff2011}. Short range correlations in the output speckle pattern can also be leveraged to enable numerical image reconstruction by means of either iterative phase retrieval or cross-correlation based algorithms \cite{Bertolotti2012,Katz2014,Newman2014a}. Alternatively, single pixel imaging techniques extract information through sequential variation of the illumination basis in combination with spatial integration of the output speckle, allowing the initial image to be rebuilt in terms of its constituent spatial modes \cite{Sun2013a,Edgar2019}. 	

As a consequence of the inherent randomisation of an image caused by transmission through a scattering medium, a computational step is always required to obtain a final image. Although good imaging results have been reported in a variety of experimental setups, the quality of such computational images can frequently be algorithm dependent, making comparison and benchmarking more difficult, an issue also encountered with other imaging modalities \cite{Verveer1999,Ram2006b,Chakrova2016}. Traditionally reported metrics of imaging performance, such as the spatial resolution or fidelity of the output image \cite{Fellgett2007}, critically do not distinguish between the detrimental effects of scattering in the medium and data post-processing.  Whilst the former is fundamental, the latter can in principle be improved through better algorithm design. It is therefore natural to ask what fundamental limitations are imposed by  transmission through a scattering medium, a problem which we consider in this work.  \revision{Notably, this limit is closely related to the degree of control achievable in wavefront shaping experiments \cite{Vellekoop2007, Horstmeyer2015} in scattering environments and thus constrains the set of realisable wave fields \cite{Hsu2017}.} Considering that the input and transmitted images are highly dissimilar, conventional imaging quality metrics are notably unsuitable to address this question. \revision{Instead we adopt an information based perspective, which has a rich history in optical imaging \cite{Fellgett2007,Linfoot1955,Linfoot1956,Black1957,Cox1986a,Huck1996,Miller2000}, whereby the output speckle pattern is treated as a message from which we extract information about the scene of interest. }

 The greater suitability of information based metrics to quantify transmission through disordered media has motivated a number of related studies \cite{Simon2001, Foschini1998, Staring2004,Andrews2001}. For instance, it has been shown that interference effects, which are prevalent in scattering environments, affects the rate of information transmission between antenna arrays \cite{Moustakas2000}. Information-theoretic metrics, such as the channel capacity of an information channel have furthermore been directly related to, and shown to decrease as a result of, mesoscopic correlations between scattered waves \cite{Skipetrov2003}. 
The mutual information between reflected and transmitted speckle images has also been recently investigated \cite{Fayard2018}.
These studies, however, are typically either  limited to the dispersive regime or require \emph{a priori} statistical knowledge of the scattering properties of the medium. Moreover, focus has generally been restricted to either wireless communications or transmission through wires and previous results are thus less applicable to an imaging context.  In this work we therefore ask the question of whether there exist fundamental system and algorithm independent bounds on how much information encoded as images can be transmitted through a complex scattering medium. Making no assumptions on the nature of the scattering medium, other than to restrict to statistical ensembles with a given mean transmittance, we show that the answer to this question is in the affirmative. We derive and discuss these bounds.  \revision{Our analysis centres around Shannon information \cite{Abramson1963} based metrics, such as mutual information and channel capacity, since these provide a direct characterisation of global information content of an image and the limits imposed by a scattering medium. Fisher information, although related \cite{Stam1959,Frieden1998a}, is less germane to the question at hand since it pertains to the task of statistical estimation and image reconstruction \cite{Scharf1991,Kay1993} as opposed to information transmission.} 
The structure of this article is therefore as follows. We begin in Section~\ref{sec:info} by formalising the information theoretic treatment of imaging through scattering media, before performing a Monte-Carlo based study of the statistical properties of some common statistical ensembles describing scattering in complex media in Section~\ref{sec:stats}. Derivation and discussion of universal ensemble independent bounds on the mean and variance of our information-theoretic metrics is given in Section~\ref{sec:bounds} where comparison to  numerical results is also given. Finally, our conclusions are given in Section~\ref{sec:conclusions}.

\section{Information in imaging through scattering media}\label{sec:info}
Before it is possible to quantify the effect of transmission through a scattering medium, it is first necessary to formalise the information content of the starting image. To do so we begin by noting that the number of degrees of freedom $N$ of an optical image are, in general, limited. For a digital image these degrees of freedom naturally correspond to the number of pixels present, whilst for analog images the limit can derive from the finite bandwidth with which the image is generated or recorded (i.e. the spatial resolution) \cite{Cox1986a}. Although finite, the degrees of freedom of an image can nevertheless be used to encode information, such that an image can be considered as a single symbol from an information source with a source alphabet of $\mathcal{S} = \{ S_1,S_2,\ldots,S_N\}$ \cite{Abramson1963}. As a concrete example, consider a digital image comprising of $N$ pixels with  a fixed total power corresponding to a single photon. The $N$ symbols can be encoded onto the position of the photon, whereby $S_j$ then corresponds to the photon being registered on the $j$th pixel of the image. The information encoded in the photon's position can then be quantified using the Shannon entropy, $H(\mathcal{S})  = -\sum_{j=1}^N p(S_j) \log p(S_j)$, where $p(S_j)$ is the probability that the photon is observed in the $j$th image pixel. The total information encoded in an image composed of $n$ photons is thus $n H(\mathcal{S})$. Note, that for ease of notation we shall use $p(\cdots)$ throughout this work to denote different probability distributions, where the associated relevant random variable will be apparent from the argument.

Although it is natural to consider each pixel of an image as an individual degree of freedom, it is equally legitimate to instead consider utilising different extended spatial modes, drawn from a complete orthonormal set of basis functions, to encode information. \revision{The basis functions must  capture the $N$ degrees of freedom present in the image.} One possible choice of such basis functions is, for example, the Hadamard functions as are frequently used in single pixel imaging \cite{Edgar2019}. With this interpretation, an arbitrary input image field can be represented as a superposition of spatial modes with associated mode coefficients $\mathbf{a} = [a_1,\ldots,a_N]$. In turn, the probability that a photon, chosen at random from all photons which make up the input image, is in the $j$th mode is given by $p(S_j) \triangleq p_j ={|a_j|^2}/{\sum_{k=1}^N |a_k|^2}$, where we define the shorthand notation $p_j$ for convenience. Naturally, each pixel can also be considered a spatial mode, whereby $|a_j|^2$ is the intensity of each image pixel.

An image field incident upon a scattering medium generates both a reflected and transmitted field, both of which can also be represented as a superposition of spatial modes. Accordingly, the effect of a medium on the incident image can  be described using the scattering matrix $\mathbb{S}$ of the medium, viz. 
\begin{align}
\left[\begin{array}{c}
\mathbf{b} \\ \mathbf{c}
\end{array}\right]  = \mathbb{S} \left[\begin{array}{c}\mathbf{a} \\ \mathbf{0}\end{array}\right]\label{eq:scatmat}
\end{align}
where $\mathbf{b} = [b_1,\ldots,b_N]$ ($\mathbf{c} = [c_1,\ldots,c_N]$) are vectors of the mode coefficients of the reflected (transmitted) fields. Note that we assume the number of input and output modes are equal for simplicity. For a lossless system we can express the scattering matrix using the polar decomposition \cite{Mello1988b} 
\begin{align}
\mathbb{S} =
\left[\begin{array}{cc}\mathbb{V} & \mathbb{O}\\
\mathbb{O} & \mathbb{U} \end{array}\right]
\left[\begin{array}{cc}  -\sqrt{1-\boldsymbol{\tau}} &\sqrt{\boldsymbol{\tau}} \\
\sqrt{\boldsymbol{\tau}} & \sqrt{1-\boldsymbol{\tau}}  \end{array}\right]\left[\begin{array}{cc} \mathbb{V}' & \mathbb{O}\\\mathbb{O} & \mathbb{U}' \end{array}\right] \label{eq:Spoldecomp},
\end{align}
where $\mathbb{O}$ is the null matrix, $\mathbb{U}$, $\mathbb{U}'$, $\mathbb{V}$ and $\mathbb{V}'$ are unitary matrices of singular vectors, and $\boldsymbol{\tau} = [\tau_1,\ldots,\tau_N]$ is a diagonal matrix containing the transmission eigenvalues. To simplify the analysis we henceforth assume the input and output modes correspond to the singular basis of the medium such that $\mathbb{U}$, $\mathbb{U}'$, $\mathbb{V}$ and $\mathbb{V}'$ in \eqref{eq:Spoldecomp} can be replaced by the identity matrix.
Since this change of basis is performed using a unitary transformation \revision{and is hence fully reversible, no spatial information is lost \cite{Zurek1990}}. Within this framework, in this work we consider three scenarios, namely measuring in i) transmission, ii) reflection or iii) both. Specifically, when measuring in transmission we input modes $\mathbf{a}$ and measure the transmitted intensities of each mode i.e. $|c_j|^2$. The output alphabet contains symbols denoted $T_1,T_2,\ldots,T_N$,  corresponding to the $N$ transmission modes. In measuring all $|c_j|^2$ ($j = 1,\ldots N$), however, we also learn about how much energy is in the aggregate of the reflected modes ($\sum_{j=1}^N |b_j|^2$) since by conservation of energy $\sum_{j=1}^N |a_j|^2 = \sum_{j=1}^N |b_j|^2 + \sum_{j=1}^N |c_j|^2$. We denote this additional possible output symbol by $T_{N+1}$, such that the complete output alphabet is $\mathcal{T} = \{ T_1,\ldots,T_N,T_{N+1}\}$. Reflection based measurements are similar albeit we now measure the reflected mode intensities $|b_j|^2$. Again through this measurement we also learn about the total transmitted intensity $\sum_{j=1}^N |c_j|^2$, such that the output alphabet is $\mathcal{R} = \{R_1,\ldots,R_N,R_{N+1}\}$. When measurements are made in both reflection and transmission the corresponding alphabet of the output is $\mathcal{U}= \{R_1,\ldots,R_N,T_1,\ldots,T_N \}$. Note that in this latter case the aggregate output symbols (i.e. the $(N+1)$th outputs) are omitted. 

Thus far it has been implicitly assumed that the basis of spatial modes used to express the input and output images is complete, \revision{in the sense that arbitrary input and output image fields, with $N$ degrees of freedom can be represented. Taking a basis of  angular channels (i.e. the Fourier domain is discretised) for concreteness}, this implies that all spatial frequencies of the original image are incident onto the scattering medium, and similarly that all output spatial frequencies are collected. The former is easily achieved by matching the numerical aperture (NA) of the illumination optics to the spatial bandwidth of the initial image. Use of finite NA collection optics, however, means that light that is scattered out of the  medium at large angles is not measured and thus the detection basis is not complete as assumed. This scenario can be approached using filtered scattering matrices, as is detailed further in Ref.~\cite{Goetschy2013}. Alternatively, the input and output bases can be expanded so as to include all possible angular modes and the undetected output modes instead incorporated into the aggregate channels described above. For example, when measuring in transmission with a finite NA lens, the aggregate channel $T_{N+1}$ would include all of the reflected modes in addition to the angular modes lying outside the NA of the collection optics. The latter is preferable from an informatic standpoint since it is more apparent where information is lost in the system.

The quality of information transmission through a scattering medium can be quantified using the mutual information per photon between the measured output mode intensities (including the aggregate modes for Cases i and ii) and the original image, defined as
\begin{align}
I_\mathcal{N} &= I(\mathcal{S};\mathcal{N}) = H(\mathcal{N}) - H(\mathcal{N}|\mathcal{S})\label{eq:Idef}
\end{align}
where $ \mathcal{N} = \mathcal{T}$, $\mathcal{R}$ or $\mathcal{U}$ for Cases i--iii respectively and 
\begin{align}
H(\mathcal{N}|\mathcal{S}) = -\sum_{N_j \in \mathcal{N}} \sum_{S_k \in \mathcal{S}} p(N_j,S_k) \log p(N_j|S_k) 
\end{align}
is the conditional entropy, or equivocation, of $\mathcal{N}$ given $\mathcal{S}$. Noting $p(N_j,S_k) = p(N_j|S_k)p_k$ and $p(N_j) = \sum_{k=1}^N p(N_j,S_k) $, the mutual information can be calculated from the source probabilities $p_k$ and the set of conditional probabilities $p(N_j|S_k)$. The latter can be found from the scattering matrix $\mathbb{S}$. From \eqref{eq:scatmat} and \eqref{eq:Spoldecomp} we have $\mathbf{b} = - \sqrt{1-\boldsymbol{\tau}}\cdot \mathbf{a}$, which implies
\begin{align}
|b_j|^2 = \sum_{k=1}^N \rho_k \delta_{jk}|a_k|^2\label{eq:bj2}
\end{align}
where $\rho_j = 1-\tau_j$ is the reflectance of the $j$th eigenmode of the scattering medium and $\delta_{jk}$ is the Kronecker delta. Upon normalising  by the total incident intensity and comparing to the law of total probability we can make the association  $p(R_j|S_k) = \rho_{k}\delta_{jk}$ for $j = 1,\ldots N$. Similarly it follows that $p(T_j|S_k) = \tau_{k}\delta_{jk}$ for $j = 1,\ldots,N$. These probabilities are sufficient when considering Case iii ($\mathcal{N}= \mathcal{U}$), however Cases i and ii require the further probabilities $p(T_{N+1}|S_k) = 1-\tau_k$ and $p(R_{N+1}|S_k) = \tau_k$ respectively, which follow by summing \eqref{eq:bj2} (and the equivalent expression for $|c_j|^2$) over $j$. Physically this embodies the fact that the transmittance (reflectance) of a medium describes the fraction of photons that are transmitted (reflected). Substituting these probabilities into \eqref{eq:Idef} we find
\begin{align}
I_{\mathcal{N}} &= H(\mathcal{S}) +  \Lambda_{\mathcal{N}}\sum_{j=1}^N P_j^{{\mathcal{N}}}\log P^{{\mathcal{N}}}_j\label{eq:IN},
\end{align} 
where we have defined $\Lambda_{\mathcal{N}} = \sum_{k=1}^N \eta_{k}^{\mathcal{N}} \,p_k$, $\eta_{k}^{\mathcal{T}} = \rho_k$, $\eta_{k}^\mathcal{R} = \tau_k$, $\eta_{k}^\mathcal{U} = 0$ and the probabilities
\begin{align}
P^{\mathcal{N}}_j = \eta_{j}^\mathcal{N} p_j / \Lambda_\mathcal{N}.
\end{align}
The first thing to note from \eqref{eq:IN} is that by measuring the intensity in all possible output modes ($\mathcal{N} = \mathcal{U}$) we are able to extract all the information contained in the original image, as would be intuitively expected given the system is assumed to be lossless. If, however, we do not measure the reflected (transmitted) modes as in Case i (Case~ii) we lose an amount of information equal to $-\Lambda_\mathcal{N} \sum_{j=1}^N P_j^{\mathcal{N}} \log P_j^{\mathcal{N}}$. Physically, $P_j^{\mathcal{N}}$ represents the probability that a photon, taken from \emph{only} those modes that are not directly measured, is in the $j$th spatial mode. Accordingly $-\sum_{j=1}^N P_j^{\mathcal{N}} \log P_j^{\mathcal{N}}$ can be interpreted as the Shannon entropy contained in only the reflected (transmitted) modes, however, when considering all possible output modes this information must be weighted by the relative fraction of energy carried by the reflected (transmitted) waves, i.e. by $\Lambda_\mathcal{N}$. 

It is well known that Shannon entropy is maximised when the probability of each underlying state is equally likely \cite{Abramson1963}. For a system with a given transmittance, it is therefore evident from \eqref{eq:IN} that the information lost is greatest when the expected energy in each of the lossy modes (i.e. the reflected or transmitted modes for Case i and ii respectively) is equal, specifically $P_j^{\mathcal{N}} = 1/N$. If the source entropy is maximised $p_j = 1/N$, this corresponds to a uniform eigenvalue spectrum. More generally, however, the information loss from transmission through a scattering medium is dependent on both the input image (due to the $p_j$ dependence) and the spectrum of transmission eigenvalues. The channel capacity of the scattering medium, which describes the maximum information that can be transmitted through the medium over the space of all input images, offers a more general, source independent method to quantify information loss. Specifically, for a fixed scattering medium, the channel capacity can be found by maximising the mutual information $I_{\mathcal{N}}$ with respect to the input probabilities $p_j$, i.e. $C_{\mathcal{N}} = \sup_{\{p_j\}} I_{\mathcal{N}}$.
Using the standard result for maximum Shannon entropy mentioned above, it follows immediately that $C_{\mathcal{U}} = \log N$  \cite{Abramson1963}. To determine $C_{\mathcal{T}}$ and $C_{\mathcal{R}}$, we must however maximise the mutual information, subject to the constraint $\sum_{j=1}^N p_j = 1$, explicitly using the method of Lagrange multipliers. So doing yields the result that ${C}_{\mathcal{N}} = - \log \widetilde{p}_k + \eta_k^{\mathcal{N}} \log \widetilde{P}_k^{\mathcal{N}}$, where we use the tilde notation to denote optimal quantities, which must hold for all $k=1,\ldots,N$. Using the definition for $P_j^{\mathcal{N}}$ to replace $\widetilde{p}_k$ in this relation yields, upon rearrangement,
\begin{align}
\widetilde{P}_k^{\mathcal{N}} = \left(\frac{\Lambda_{\mathcal{N}} \exp[{C}_{\mathcal{N}}]}{\eta_k^{\mathcal{N}}}\right)^{1/(\eta_k^{\mathcal{N}}-1)}\label{eq:Pkopt}.
\end{align}
Summing over $k$ gives the transcendental equation
\begin{align}
1 = \sum_{k=1}^N \left(\frac{\Lambda_{\mathcal{N}} \exp[C_{\mathcal{N}}]}{\eta_k^{\mathcal{N}}}\right)^{1/(\eta_k^{\mathcal{N}}-1)},\label{eq:Ctranseq}
\end{align}
which with knowledge of all transmission eigenvalues can be solved  numerically to find $\Lambda_{\mathcal{N}} \exp[{C}_{\mathcal{N}}]$. The optimal $\widetilde{P}_k^{\mathcal{N}}$ then follow from \eqref{eq:Pkopt}. In turn the input probabilities that maximise the mutual information and achieve the channel capacity follow according to
\begin{align}
\widetilde{p}_k^{\mathcal{N}} = \frac{\widetilde{P}_k^{\mathcal{N}}}{\eta_k^{\mathcal{N}} }\Bigg/\sum_{j=1}^N \frac{\widetilde{P}_j^{\mathcal{N}}}{\eta_j^{\mathcal{N}}}
\end{align}
which further allows $\Lambda_\mathcal{N}$ and $C_\mathcal{N}$ to be calculated individually. \eqref{eq:Ctranseq} is formally equivalent to the condition found for the channel capacity of a reduced information channel \cite{Abramson1963}. Although still dependent on the precise details of the spectrum of transmission (or reflection) eigenvalues, i.e. $\eta_k^{\mathcal{N}}$, in Appendix~\ref{app:convex} we show that both $C_{\mathcal{N}}$ and $I_{\mathcal{N}}$ are decreasing functions of $\eta_k^{\mathcal{N}}$. Accordingly we find that modes with larger $\eta_k^{\mathcal{N}}$ individually contribute more significantly to the total information content of the measured signal. It should also be noted that in the derivation above it has been  assumed that $0<\eta_k^{\mathcal{N}} <1$. If $\eta_k^{\mathcal{N}} = 0$ or $1$ for some $k$ additional care must be taken in the maximisation. Illustration of how such cases can be approached is given in Appendix~\ref{app:bimodal} where we derive the channel capacity for the extreme case of a bimodal information channel, i.e. one for which $\eta_k^{\mathcal{N}}$ are Bernoulli random variables. Specifically, we show that when $K$ elements of the vector $\boldeta = [\eta_1^{\mathcal{N}},\ldots,\eta_N^{\mathcal{N}} ]$ are zero, corresponding to  so-called open channels \cite{Rotter2017}, ($N-K$ elements are hence unity) the channel capacity is given by $C_{\mathcal{N}} = \log[K + 1 - \delta_{KN}]$.

\section{Statistical properties of informatic metrics}\label{sec:stats}
\begin{figure}[b!]
	\begin{center}
		\includegraphics[width=\columnwidth]{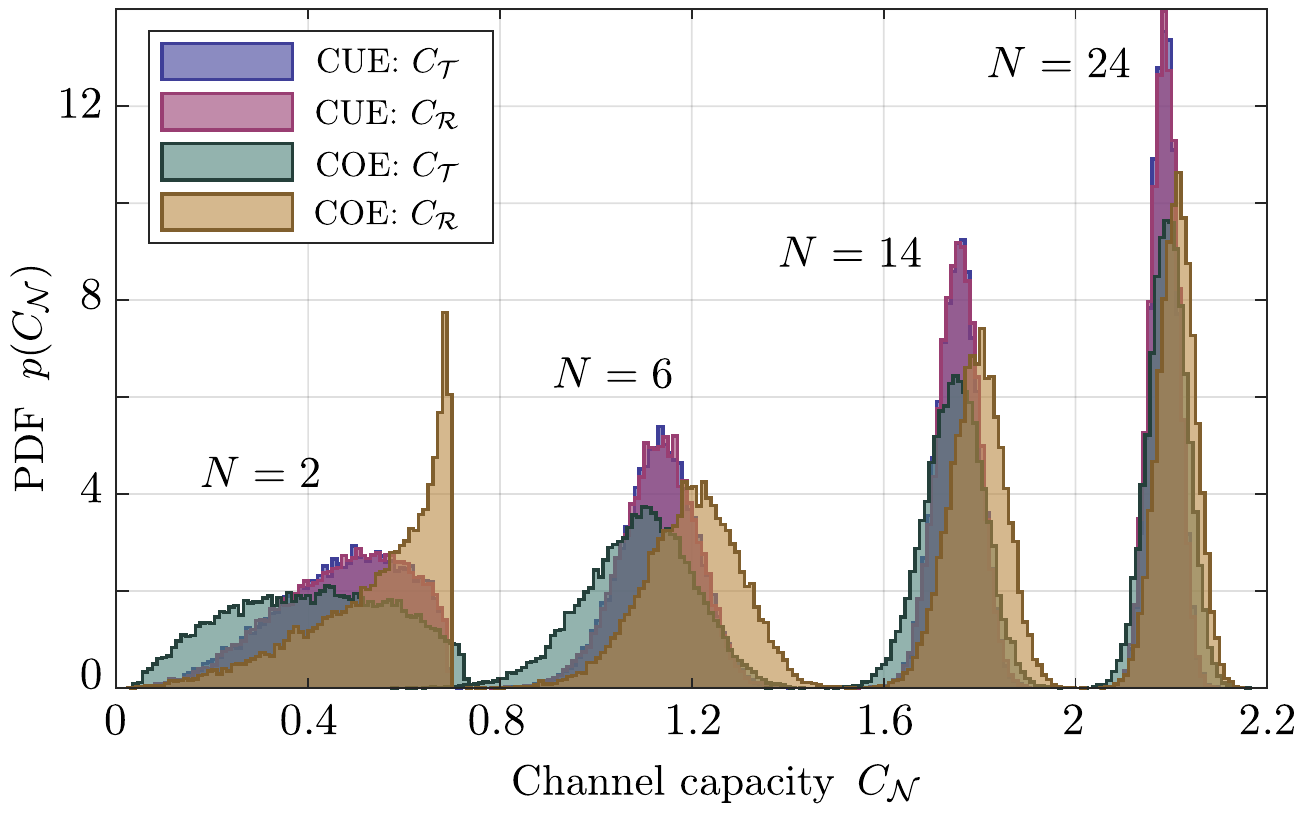}%_C_selected_N_all_ensemblesS
		\caption{Histograms of channel capacities $C_{\mathcal{N}}$ for imaging in transmission and reflection, for differing mode numbers $N$, calculated using $2\times 10^4$ realisations of transmission eigenvalues drawn from the CUE and COE. \label{fig:hist}}
	\end{center}
\end{figure}

The transmission properties of scattering media are in general random and thus the mutual information upon transmission of a known image and the channel capacity of the medium differ case by case. If the scattering matrix of a given medium is known,  individual results can nevertheless be calculated. The stochastic variability of these information-theoretic quantities is however intrinsically linked to the underlying probability distribution function (PDF) of the transmission eigenvalues, $p(\boldsymbol{\tau})$, and thus differs depending on the physics at hand. Chaotic behaviour, for instance, can arise in systems where waves are scattered from structures whose typical dimensions are large relative to the wavelength of the scattered waves, whereby ray dynamics is dominant \cite{Doron1991}. Random matrix theory is known to provide a good statistical description of the scattering matrices of such systems \cite{Beenakker1997,Rotter2017} through use of Dyson's circular ensembles \cite{Dyson1962}. Random disordered systems, in which incident flux is on average equally distributed among all possible outgoing channels, are however better described using the theory of Dorokhov \cite{Dorokhov1982}, and Mello, Pereyra and Kumar \cite{Mello1988} (DMPK), which describes evolution of $p(\boldsymbol{\tau})$ with medium thickness using a Fokker-Planck equation. In each of these cases, and indeed more generally, the precise form of the PDF governing the transmission eigenvalues (and hence $I_{\mathcal{N}}$ and $C_{\mathcal{N}}$) is dependent on the number of transmission modes $N$ \cite{Beenakker1997}. To illustrate this point, in Fig.~\ref{fig:hist} we plot PDFs of the channel capacities $C_\mathcal{T}$ and $C_{\mathcal{R}}$, calculated using $2\times 10^4$ realisations of scattering matrices sampled from the circular unitary (CUE) and orthogonal ensembles (COE), for different $N$. 
Scattering matrices drawn from the CUE are only constrained so far as to ensure $\mathbb{S}$ is unitary, such that the PDFs found for transmission (blue bars) and reflection (purple) are identical (within statistical fluctuation). The further constraint of time reversal symmetry is however imposed on scattering matrices sampled from the COE. By virtue of the resulting coherent back scattering, in which time reversed scattering trajectories constructively interfere \cite{Beenakker1997}, it is seen that the channel capacity for measurements made in reflection (yellow bars) is on average larger than that for transmission measurements (green). Differences in $p(C_{\mathcal{R}})$  and $p(C_{\mathcal{T}})$ are particularly marked for low $N$, however, for larger numbers of modes the PDFs both converge to a normal distribution, suggestive of a central limit theorem (CLT). Similar asymptotic behaviour is also seen for the CUE and DMPK ensemble and for PDFs of the mutual information (not shown). 

Each term in the summation of \eqref{eq:IN} (and the equivalent for $C_{\mathcal{N}}$) can be considered as different random variables, however, existence of CLTs (for $\mathcal{N}\neq \mathcal{U}$) for large $N$ is non-trivial since non-zero correlations between the transmission eigenvalues violate the usual independent random variable approximations required for conventional CLTs to hold \cite{Leon-Garcia1994}. It has been shown that any linear statistic of the transmission eigenvalues \cite{Kamien1988} does still obey a CLT even in the presence of eigenvalue correlations, however it is important to note that $I_{\mathcal{N}}$ and $C_{\mathcal{N}}$ are not linear statistics of $\tau_j$. This observation is in contrast to previous works, e.g. \cite{Staring2004}, and is a result of the aggregate symbols $R_{N+1}$ and $T_{N+1}$ which, by definition, depend on all eigenvalues. This non-linear dependance also implies that the mean properties of $I_{\mathcal{N}}$ and $C_{\mathcal{N}}$ are not wholly determined by the average eigenvalue spectrum, since higher order statistical moments can play an important role. CLTs for a restricted class of multi-linear statistics have been demonstrated \cite{Lytova2008}, however, in the current context the CLT can be most easily justified through extension of the rationale of Ref.~\cite{Politzer1989} for linear statistics. This is detailed in Appendix~\ref{app:CLT}. Existence of such CLTs is restricted to the same cases as described in Ref.~\cite{Kamien1988} and do not exist for all possible ensembles.

\begin{figure}[t!]
	\begin{center}
		\includegraphics[width=\columnwidth]{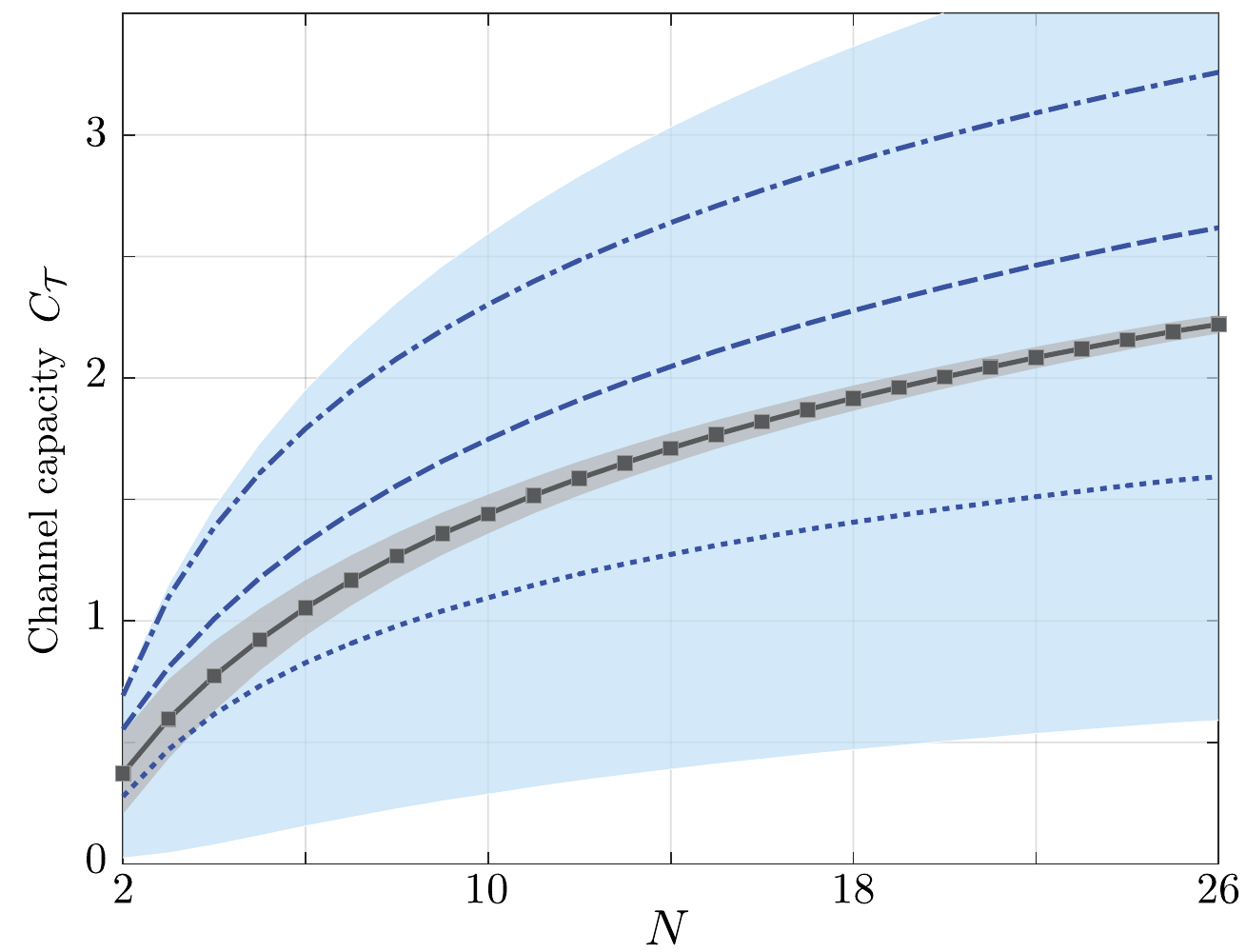}
		\caption{Mean channel capacity $\bar{C}_{\mathcal{T}}$ of a random medium vs number of modes $N$ calculated using $2\times 10^4$ transmission eigenvalues drawn from a COE (dark gray line with square markers). Dotted line represents lower bound of $(1-\bar{\eta})\log N$, dashed line shows upper bound described by \eqref{eq:C_GUB} whereas dot-dashed line shows upper bound of $\log N$. Shaded gray (blue) area depicts the band defined by $\bar{C}_{\mathcal{T}} \pm \sigma_{\bar{C}_{\mathcal{T}}}$ calculated using Monte Carlo results (minimum of \eqref{eq:Cvar_bound1} and \eqref{eq:Cvar_bound2}).}\label{fig:COE_vs_N}
	\end{center}
\end{figure}

\begin{figure}[t!]
	\begin{center}
		\includegraphics[width=\columnwidth]{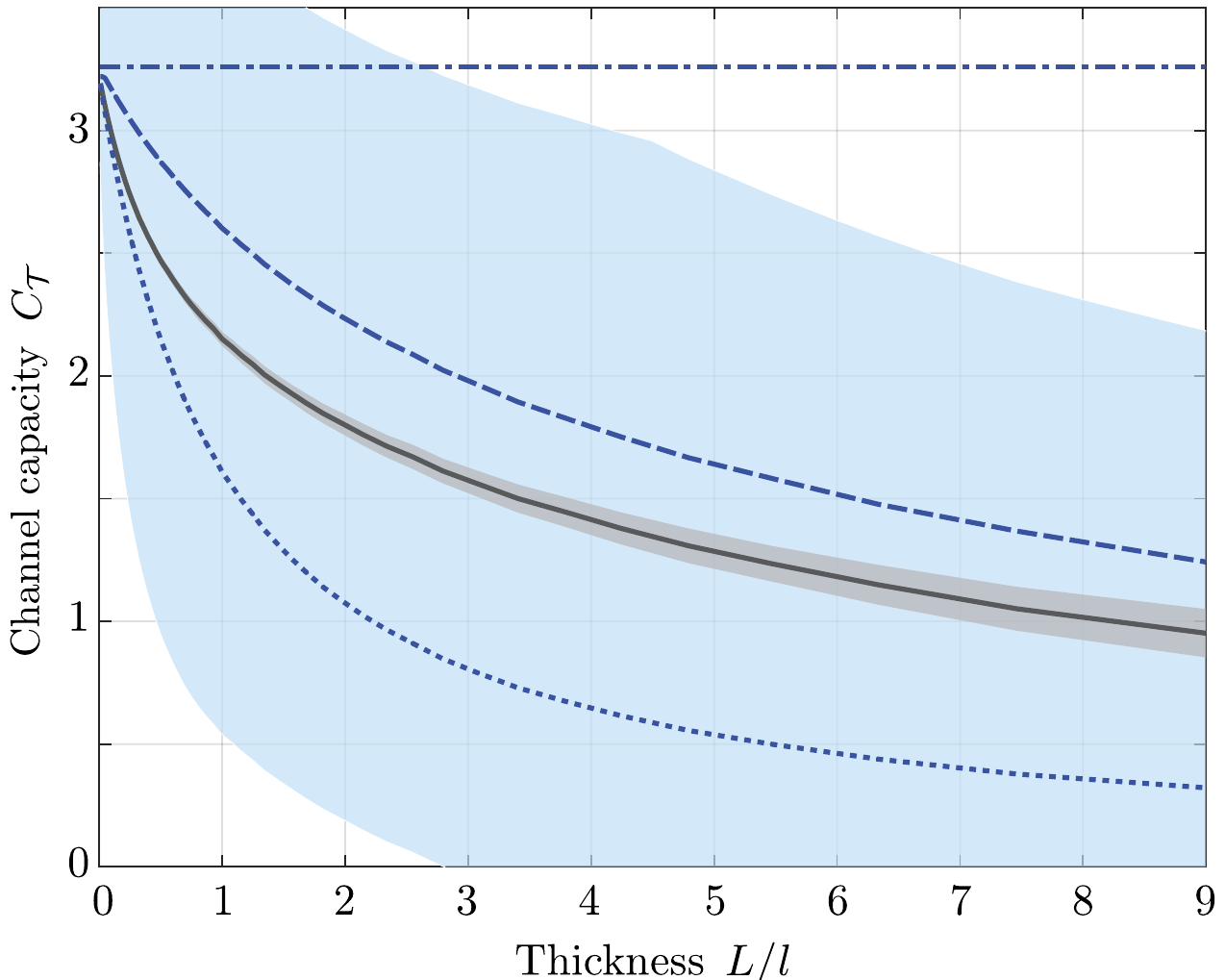}
		\caption{As Fig.~\ref{fig:COE_vs_N}, except transmission eigenvalues are drawn from the DMPK ensemble, and channel capacities are plotted as a function of the scattering medium thickness $L$ relative to the mean free path $l$.}\label{fig:DMPK_vs_s}
	\end{center}
\end{figure}

Full statistical parametrisation of the channel capacity and mutual information can give deep insights, however, from a practical point of view the values expected on average, and the degree to which they vary between media of the same statistical class, are more convenient. Accordingly, we here consider $\bar{C}_{\mathcal{N}} =  E_{\boldsymbol{\eta}}[C_{\mathcal{N}}]$  and $\sigma_{C_{\mathcal{N}}}^2 = E_{\boldsymbol{\eta}}[C_{\mathcal{N}}^2] - \bar{C}_{\mathcal{N}}^2$ (and analogous quantities for $I_{\mathcal{N}}$), where $E_{\boldsymbol{\eta}}[\cdots]$ denotes the statistical expectation over the ensemble of possible ${\boldsymbol{\eta}}$ (or equivalently ${\boldsymbol{\tau}}$). Given the CLTs discussed above, for large $N$ these parameters can be sufficient to uniquely describe the full PDF. Figure~\ref{fig:COE_vs_N} shows the dependence of the average channel capacity for transmission measurements ($\bar{C}_\mathcal{T}$) as a function of $N$ (dark gray line with square markers) when the scattering matrix is drawn from the COE as calculated using Monte-Carlo simulations. The shaded gray area, moreover, depicts the corresponding band defined by $\bar{C}_{\mathcal{T}} \pm \sigma_{\bar{C}_{\mathcal{T}}}$. Average channel capacity is seen to increase sub-linearly as mode number increases in contrast to other geometries in which a linear increase has been found \cite{Simon2001,Moustakas2000}. Furthermore Fig.~\ref{fig:DMPK_vs_s} shows  $\bar{C}_{\mathcal{T}}$ for scattering matrices drawn from a DMPK ensemble for  disordered media of varying thicknesses $L$ (measured in mean free paths $l$). Scattering matrices in this case were generated using the technique detailed in Ref.~\cite{Li2019}. It is evident that channel capacity in transmission decreases as the mean transmittance of each eigenchannel (which are equal for all eigenchannels within the DMPK model) decreases since $\bar{\tau}\sim(1+L/l)^{-1}$ \cite{Mello1991}.

\section{Universal bounds on mean information}\label{sec:bounds}
Although numerical results, as shown in Figs.~\ref{fig:hist}--\ref{fig:DMPK_vs_s}, are insightful, exact analytic results are preferable. The complexity of the PDFs governing $\boldsymbol{\tau}$ however preclude determination of exact analytic results. Moreover, ensemble specific results are somewhat restrictive and not applicable to different classes of scattering media. As such we instead now consider derivation of ensemble independent informatic bounds. For measurements made in both reflection and transmission (Case iii) our analysis is particularly simple, since all information is retrieved such that  $\bar{{C}}_{\mathcal{U}} =\log N$, $\bar{{I}}_{\mathcal{U}} = H(\mathcal{S})$ and $\sigma_{{C}_{\mathcal{U}}}^2 = \sigma_{{I}_{\mathcal{U}}}^2 = 0$. For Cases i and ii, whilst it is immediately obvious that regardless of the underlying statistics, $C_{\mathcal{N}} \leq \log N$ and $I_{\mathcal{N}} \leq H(\mathcal{S})$, tighter upper bounds, parametrised only by the mean reflectance or transmittance respectively, can be derived. To do so relies upon the observation that both $C_{\mathcal{N}}$ and $I_{\mathcal{N}}$ are convex functions with respect to $\eta_k^{\mathcal{N}}$, which we prove in Appendix~\ref{app:convex}. Since the analysis is identical for $\mathcal{N} = \mathcal{R}$ and $\mathcal{T}$ we temporarily drop the $\mathcal{N}$ superscripts on $\eta_j$ for clarity.

Convexity of both the channel capacity and mutual information means that lower bounds for their expectations immediately follow from Jensen's inequality \cite{Boyd2010}, namely $C_{\mathcal{N}}(\bar{\boldsymbol{\eta}}) \leq \bar{C}_{\mathcal{N}}$ and $I_{\mathcal{N}}(\bar{\boldsymbol{\eta}}) \leq \bar{I}_{\mathcal{N}}$ where $\bar{\boldsymbol{\eta}} = E_{\boldsymbol{\eta}}[{\boldsymbol{\eta}}]$. For a balanced channel for which all mean eigenvalues are equal ($\bar{\eta}_j = \bar{\eta}$ for all $j$) these lower bounds take the simple form $(1-\bar{\eta}) \log N  \leq \bar{C}_{\mathcal{N}}$ and $(1-\bar{\eta}) H(\mathcal{S})  \leq \bar{I}_{\mathcal{N}}$. Equality is achieved for a deterministic medium with fixed transmittance.  We thus note that the channel capacity of a random scattering medium is on average larger than that of a deterministic channel. 

Derivation of an upper bound for $\bar{C}_{\mathcal{N}}$ again invokes convexity of $C_{\mathcal{N}}$. Specifically, convexity with respect to $\eta_1$ implies that  $C_{\mathcal{N}}(\eta_1,\eta_2,\ldots) \leq (1-\eta_1)C_{\mathcal{N}}(0,\eta_2,\ldots) + \eta_1 C_{\mathcal{N}}(1,\eta_2,\ldots)$. Since $C_{\mathcal{N}}$ is convex with respect to all $\eta_j$ similar inequalities can be sequentially applied yielding
\begin{align}
\bar{C}_{\mathcal{N}}\leq \sum_{\mathbf{u} \in P} C_{\mathcal{N}}\left(\mathbf{u}\right) E_{\boldsymbol{\eta}}\Bigg[\prod_{i \in A_\mathbf{u}  } (1-\eta_i) \prod_{j \in B_\mathbf{u}} \eta_j \Bigg], \label{eq:CNbarUB}
\end{align}
where $P$ is the set of all $N$-tuples $\mathbf{u}= [u_1,u_2,\ldots,u_N]$ of $\{0,1\}$ (i.e. $u_j = 0$ or $1$ for all $j$), whereas $A_\mathbf{u}$ and $B_\mathbf{u}$ are the sets $A_\mathbf{u} = \{i | 1\leq i \leq N, u_i = 0 \}$ and $B_\mathbf{u} = \{j |1\leq j \leq N,  u_j = 1 \}$. The cardinality of the sets are $\# P = 2^N$, $\# A_\mathbf{u} = K$ and $\# B_\mathbf{u} = N-K = |\mathbf{u}|^2$. \eqref{eq:CNbarUB} gives an upper bound no worse than $\log N$ for arbitrary unconstrained ensembles as shown in Appendix~\ref{app:bound}. Letting $w(\mathbf{u}) = \prod_{i \in A_\mathbf{u} } (1-\eta_i) \prod_{j \in B_\mathbf{u}} \eta_j$ and $\bar{w}(\mathbf{u}) = E_{\boldsymbol{\eta}}[w(\mathbf{u})]$ we observe that 
\begin{align}
\sum_{\mathbf{u}\in P} w(\mathbf{u}) = \sum_{\mathbf{u}\in P} \bar{w}(\mathbf{u}) = 1.\label{eq:con1} 
\end{align}
Accordingly the upper bound in \eqref{eq:CNbarUB} represents the weighted average of the channel capacity that can be sent through bimodal information channels (i.e. an information channel for which $\eta_i = 0$ or 1 for all $i$). The weightings in the mixture, however, depend on the statistics of the transmission eigenvalues as parametrised by the set of $\bar{w}(\mathbf{u})$. A universal, i.e. ensemble independent, upper bound on the mean channel capacity can thus be found by maximising the right-hand side of \eqref{eq:CNbarUB} with respect to the expectations $\bar{w}(\mathbf{u})$. This maximisation is however performed subject to a number of constraints on $\bar{w}(\mathbf{u})$ beyond that given by \eqref{eq:con1}. Firstly we note that because $0\leq \eta_i \leq 1$ for all $i$,
the weights are themselves bounded such that $0 \leq \bar{w}(\mathbf{u}) \leq 1$. 
Furthermore for Case i (ii) we can consider the total reflectance (transmittance) given by $g(\boldsymbol{\eta}) = \sum_{i = 1}^N \eta_i$. Noting that $g_{\mathcal{N}}(\ldots,\eta_j,\ldots) = (1-\eta_j) g_{\mathcal{N}}(\ldots,0,\ldots) + \eta_j g_{\mathcal{N}}(\ldots,1,\ldots)$ it follows (similarly to above) that 
\begin{align}
\sum_{\mathbf{u}\in P}(N-K) \bar{w}(\mathbf{u}) =  \sum_{i=1}^N\bar{\eta}_i \triangleq \bar{g}_{\mathcal{N}},\label{eq:con2}
\end{align}
where $\bar{\eta}_i = E_{\boldsymbol{\eta}}[\eta_i]$ is the mean of the $i$th transmission eigenvalue. The right hand side of \eqref{eq:con2} physically corresponds to the mean total reflectance (transmittance) of the scattering medium, denoted $\bar{g}_{\mathcal{N}}$.

When considering the universal upper bound on the mean mutual information $\bar{I}_{\mathcal{N}}$, the maximisation must in general be performed numerically since the bound is strongly dependent on the source image through $p_j$. Analytic results can however be found for the universal upper bound on the channel capacity by noting that $C_{\mathcal{N}}(\mathbf{u})$ in \eqref{eq:CNbarUB} represents the channel capacity for a bimodal channel with $K$ open channels, i.e. $C_{\mathcal{N}}(\mathbf{u}) = \log[K + 1 - \delta_{KN}]$ (see Appendix~\ref{app:bimodal}). Since $C_{\mathcal{N}}(\mathbf{u})$ depends only on the number of zero elements in $\mathbf{u}$ and not on the ordering of the elements \eqref{eq:CNbarUB} can be written as 
\begin{align}
\bar{C}_{\mathcal{N}}\leq \sum_{K=0}^N \widetilde{w}_K \log[K + 1 - \delta_{KN}] \label{eq:CNbarUB2}
\end{align}
where $\widetilde{w}_K = \sum_{\mathbf{u} \in U_K }  \bar{w}(\mathbf{u})$ is the sum of the weights over the set $U_K$ of bimodal channels with $|\mathbf{u}|^2 = N-K$. For a system with total mean reflectance (transmittance) of $\bar{g}_{\mathcal{N}}$ the constraints can be similarly written $\sum_{K= 0}^N \widetilde{w}_K = 1$ and $\sum_{K= 0}^N (N-K) \widetilde{w}_K = \bar{g}_{\mathcal{N}}$. Maximisation of the right hand side of \eqref{eq:CNbarUB2} occurs when 
\begin{align}
\widetilde{w}_j = \left\{ \begin{array}{ll}
k+1 - \bar{g}_{\mathcal{N}} & \mbox{if } j = N-k\\
\bar{g}_{\mathcal{N}} - k & \mbox{if } j = N-k-1\\
0 & \mbox{otherwise}
\end{array} \right.\label{eq:Cbar_wopt}
\end{align}
where $k = \mbox{floor}[\bar{g}_{\mathcal{N}}]$. The corresponding universal upper bound on the channel capacity (see Appendix~\ref{app:capacity}) is hence $\bar{C}_{\mathcal{N}} \leq \bar{C}^{\txtpow{max}}_{\mathcal{N}}$ where
\begin{align}
\bar{C}^{\txtpow{max}}_{\mathcal{N}} &= [k+1 - \bar{g}_{\mathcal{N}}] \log[N-k+1-\delta_{k0}] \nonumber\\
&\quad\quad +
[\bar{g}_{\mathcal{N}} - k] \log[N-k]   . \label{eq:C_GUB}
\end{align}
Both the upper and lower bounds on the mean channel capacity are  shown in Figs.~\ref{fig:COE_vs_N} and \ref{fig:DMPK_vs_s} (dashed and dotted curves respectively), in addition to the weaker upper bound of $\log N$ (dot-dashed curve). Neither of these upper or lower bounds can be improved without further restricting the properties of the statistical ensembles under consideration. 

\begin{figure}[t!]
	\begin{center}
		\includegraphics[width=0.95\columnwidth]{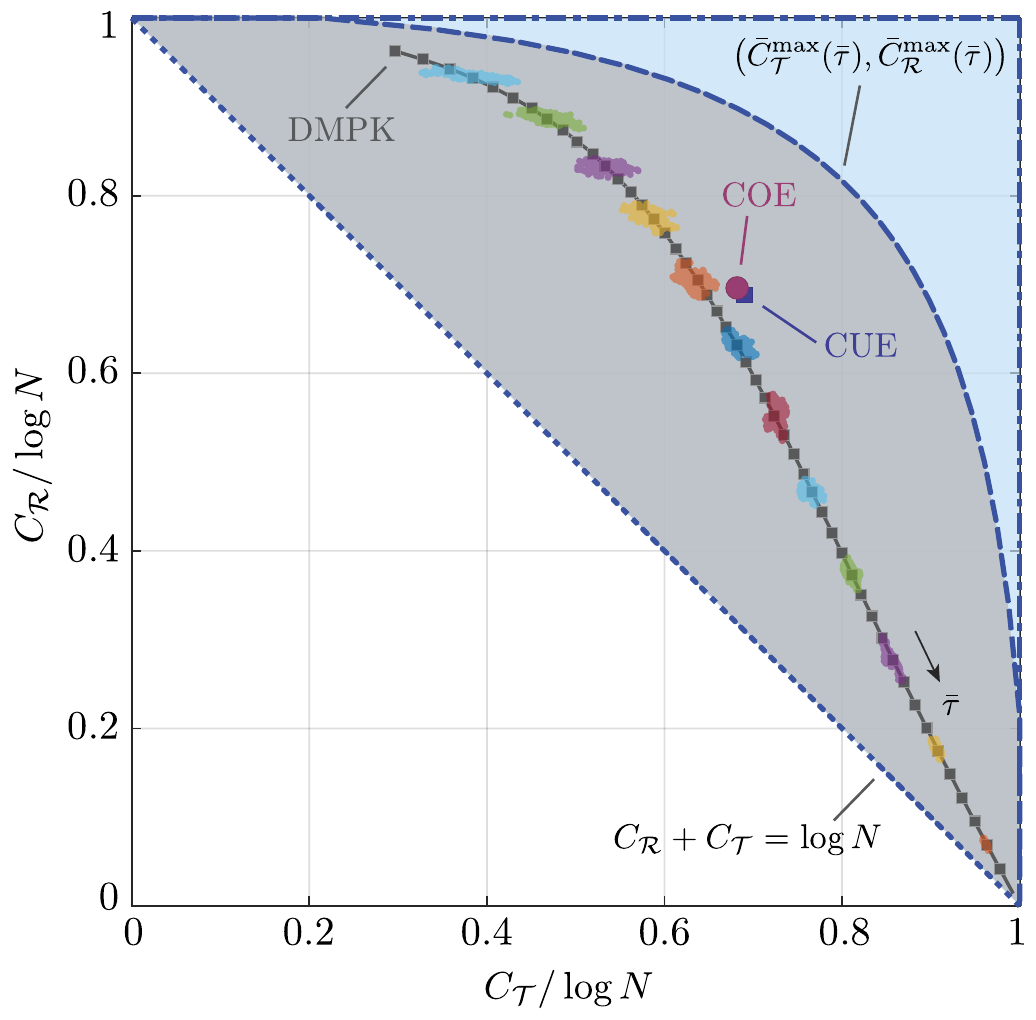}
		\caption{$(C_{\mathcal{T}},C_{\mathcal{R}})$ plane depicting allowed domains for channel capacities of individual scattering media (blue and gray shaded area combined) and ensemble averaged capacities (gray shaded area only). Bounding curves correspond to $C_{\mathcal{R}} + C_{\mathcal{T}} = \log N$ (dotted blue), $C_{\mathcal{R}} = \log N$ and $C_{\mathcal{T}} = \log N$ (dot-dashed blue curve) and parametric curve defined by $(\bar{C}_{\mathcal{T}}^{\txtpow{max}}(\bar{\tau}),\bar{C}_{\mathcal{R}}^{\txtpow{max}}(\bar{\tau}))$ (dashed blue). Positions of average capacities $(\bar{C}_{\mathcal{T}},\bar{C}_{\mathcal{R}})$ for COE (purple circle) and CUE (blue square) are shown. Solid gray curve shows the mean  $(\bar{C}_{\mathcal{T}}(\bar{\tau}),\bar{C}_{\mathcal{R}}(\bar{\tau}))$ curve for the DMPK ensemble (solid gray) found from Monte Carlo simulations.  Point clouds correspond to 250 individual realisations drawn from the DMPK ensemble with differing mean transmittance $N\bar{\tau}$.  $N=25$ was used for all ensembles.}\label{fig:CRCT_plane}
	\end{center}
\end{figure}
\begin{figure}[t!]
	\begin{center}
		\includegraphics[width=0.95\columnwidth]{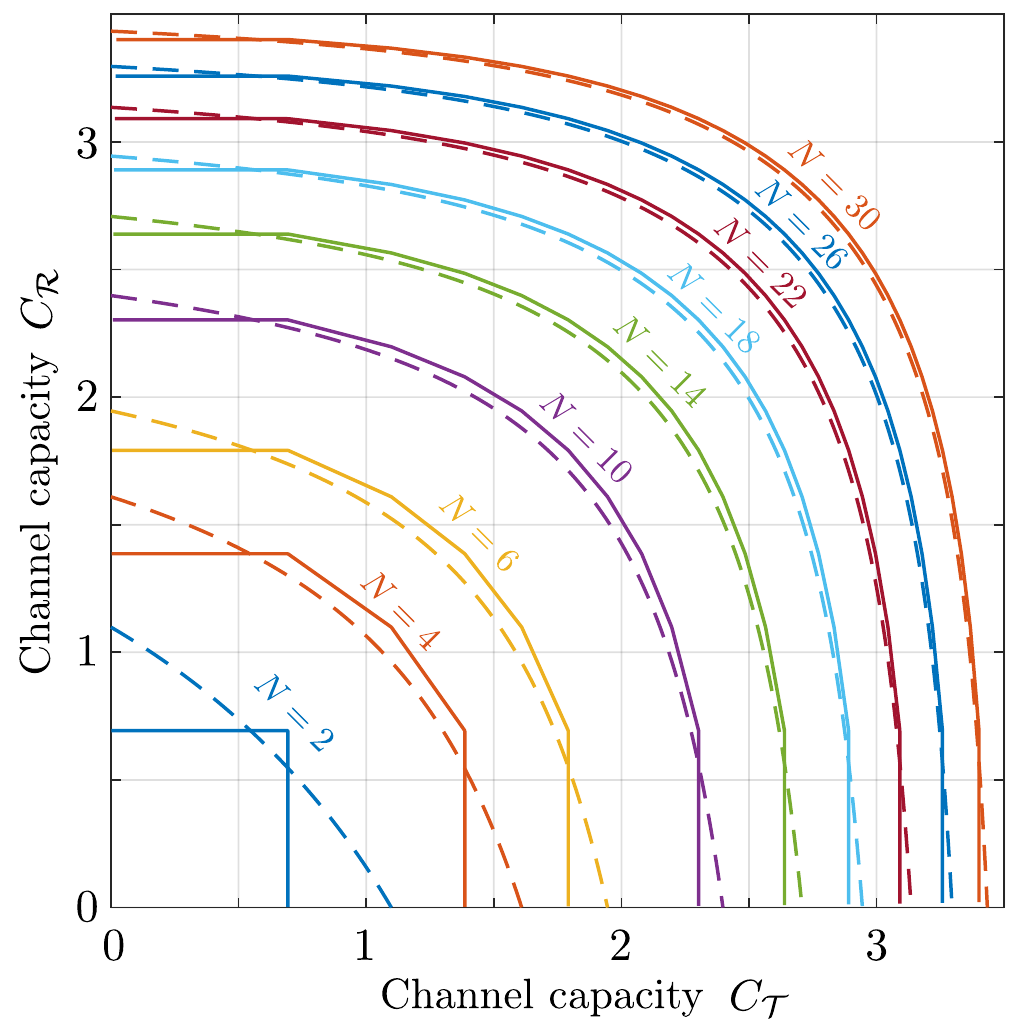}
		\caption{Comparisons of the upper bound on the channel capacity $C_{\mathcal{N}}^{\txtpow{max}}$ (solid curves) to that of a bimodal channel with independent identically Bernoulli distributed transmission eigenvalues (dashed curves) in the $(C_{\mathcal{T}},C_{\mathcal{R}})$ domain.}\label{fig:bimodal_compare}
	\end{center}
\end{figure}

We have seen above that when measurements are made in both reflection and transmission (Case iii) all information encoded in the original image can be extracted. It may hence be intuitively expected that if the mutual information (or channel capacity) for transmission measurements is larger, then the corresponding value for reflection measurements is smaller. Each pair of metrics, e.g. $(C_{\mathcal{T}},C_{\mathcal{R}})$, define an information plane as illustrated in Fig.~\ref{fig:CRCT_plane}, on which such relations can be visualised.
Through simple manipulation of \eqref{eq:IN} and application of Gibbs' inequality \cite{Abramson1963} it can be shown that $H(\mathcal{S}) \leq I_{\mathcal{R}} + I_{\mathcal{T}} \leq 2 H(\mathcal{S})$ and similarly $\log N \leq C_{\mathcal{R}} + C_{\mathcal{T}}\leq 2 \log N$. In combination with our earlier bounds these inequalities imply that a single scattering medium drawn from any ensemble with fixed mean transmittance is described by a single point lying in a triangular region of the associated information plane. Figure~\ref{fig:CRCT_plane} illustrates this permissible region (combination of the blue and gray shaded areas) when considering the $(C_{\mathcal{T}},C_{\mathcal{R}})$ plane. Note that for the case of a balanced ensemble, the $C_{\mathcal{T}}+ C_{\mathcal{R}} = \log N$ boundary (dotted blue line) corresponds to the lower bound given by Jensen's inequality found above. Channel capacities for individual realisations of scattering media drawn from the DMPK ensemble are also shown in Fig.~\ref{fig:CRCT_plane} assuming $N=25$. Distinct clusters of points are evident and correspond to differing mean transmittances ($N\bar{\tau}$) and lie along the parametric curve (solid gray curve with markers) defined by $(\bar{C}_{\mathcal{T}},\bar{C}_{\mathcal{R}})$. Individual realisations are shown for equally spaced values of $\bar{\tau}$ ranging from 0.95 to 0.15. As discussed above decreasing $\bar{\tau}$ corresponds to thicker samples. Average channel capacities for thicker samples are hence again seen to be greater in a reflection modality.  Noting that $\log N \leq \bar{C}_{\mathcal{R}} + \bar{C}_{\mathcal{T}}$ also holds, such parametric curves for the mean channel capacities of other statistical ensembles must also lie within the triangular region shown in Fig.~\ref{fig:CRCT_plane}. Bounds on $\bar{C}_{\mathcal{N}}$ derived above, however, further restrict the allowed region to that depicted by the gray shading. $(\bar{C}_{\mathcal{T}},\bar{C}_{\mathcal{R}})$ points corresponding to the COE (purple circle) and CUE (blue square) are also shown and clearly lie within this admissible region. 

Interestingly, we also find that in the asymptotic limit of large $N$, the upper bound defined by $(\bar{C}_{\mathcal{T}}^\txtpow{max},\bar{C}_{\mathcal{R}}^\txtpow{max})$ can be well approximated by the mean channel capacity found when the transmission eigenvalues are independent identically distributed Bernoulli random variables with mean of $\bar{\eta}= \bar{g}_{\mathcal{N}}/N$ namely
\begin{align}
\bar{C}_{\mathcal{N}}^b = \sum_{j=0}^N {{N}\choose{j}} \bar{\eta}^j (1-\bar{\eta})^{N-j} \log[N-j+1-\delta_{j0}]. 
\end{align}
The quality of this approximation is shown in Fig.~\ref{fig:bimodal_compare}.  At this point we also note an interesting connection with the results of Ref.~\cite{Hsu2017} in which it was demonstrated that in the diffusive regime there are at best $g_{\mathcal{T}}$ degrees of freedom when attempting to focus light through a disordered medium. In this scattering regime the eigenvalue spectrum corresponds to a bimodal distribution which is highly concentrated at both $\tau_k \approx 0$ and $\tau_k \approx 1$ \cite{Mello1988}. The degrees of freedom available to engineer the light field in a scattering medium are thus those that preserve the information about the source field. Our results show that this intuitive rule also represents a rigorous limit beyond the diffusive scattering regime. Moreover, within an imaging context, we note the aggregate channel provides additional information.

Finally we consider what universal bounds exist for the variance of the channel capacity (our discussion will be solely in terms of the channel capacity, however analogous results hold for the mutual information $I_{\mathcal{N}}$).  As with the case for the channel capacity, a deterministic scattering medium provides the trivial lower bound on the variance in which case $\sigma^2_{C_{\mathcal{N}}}  = 0$. For any bounded random variable $X$ ($0\leq X \leq 1$) the Bhatia-Davis inequality states that variance of $X$ has a maximum value of $\bar{x}(1-\bar{x}) $ when $X$ is Bernoulli distributed and where $\bar{x}$ is the mean of $X$ (or equivalently the probability that $X=1$) \cite{Bhatia2000}. Transforming this result onto the problem of determining the maximal value of $\sigma^2_{C_{\mathcal{N}}}$ we have $\sigma^2_{C_{\mathcal{N}}} \leq \bar{C}_{\mathcal{N}}(\log N - \bar{C}_{\mathcal{N}})$.  This expression is however not ensemble independent due to the dependence on the mean channel capacity. Instead the variance must be maximised subject to the inequality constraints derived above. Maximum variance is again achieved when $C_{\mathcal{N}}$ is Bernoulli distributed whereby 
\begin{align}
\!\!\sigma^2_{C_{\mathcal{N}}} \leq \left\{\begin{array}{lll} 
C(\bar{\boldsymbol{\eta}}) (\log N - C(\bar{\boldsymbol{\eta}})  ) & \mbox{if~} 2C(\bar{\boldsymbol{\eta}}) \geq \log N ,\\
\bar{C}^{\txtpow{max}}_{\mathcal{N}}(\log N - \bar{C}^{\txtpow{max}}_{\mathcal{N}}) & \mbox{if~} 2\bar{C}^{\txtpow{max}}_{\mathcal{N}} \leq  \log N ,\\
(\log N)^2/4 & \mbox{otherwise.}
\end{array}\right.  \label{eq:Cvar_bound1}
\end{align}
We note that when the CLT discussed above holds, \eqref{eq:Cvar_bound1} only gives a loose bound on the variance due to the differing nature of the Gaussian and Bernoulli distributions applicable in each case. The significant difference between the calculated variances and the limiting values is evident in Figs.~\ref{fig:COE_vs_N} and \ref{fig:DMPK_vs_s}.

An alternative upper bound on the variance of the channel capacity can however also be derived which can sometimes give slightly improved constraints in comparison to \eqref{eq:Cvar_bound1} (this accounts for the slight kink in the blue band plotted in Fig.~\ref{fig:DMPK_vs_s}). To do so we first note that $C_{\mathcal{N}}$ is a non-negative convex function with respect to $\boldeta$, such that taking the square preserves convexity \cite{Boyd2010}. Following a maximization procedure similar to that given above, albeit for $E_{\boldeta}[C_{\mathcal{N}}^2]$, gives
\begin{align}
\sigma_{C_{\mathcal{N}}}^2 \leq \widetilde{w}_\beta \log[\beta+1-\delta_{\beta N}]^2 + \widetilde{w}_\gamma \log[\gamma+1]^2  -   C_{\mathcal{N}}^2(\bar{\boldeta}), \label{eq:Cvar_bound2}
\end{align}
where $\widetilde{w}_\gamma = 1- \widetilde{w}_\beta$ and we have used the lower limit on $\bar{C}_{\mathcal{N}}$ to express the bound in an ensemble independent manner.  The explicit expressions for $\beta$, $\gamma$ and $\widetilde{w}_\beta$ are dependent on both $m = \text{min}[N-1,4]$ and $k = \text{floor}[\bar{g}_{\mathcal{N}}]$. Specifically if $k < n-m$ then $\beta = N-k$, $\gamma = \beta - 1$ and $\widetilde{w}_{\beta} = k + 1 - \bar{g}_{\mathcal{N}}$. Alternatively if $k\geq n-m$ then $\beta = m$, $\gamma = 0$ and $\widetilde{w}_{\beta} = \left(N-{\bar{g}_{\mathcal{N}}}\right)/m$. 

\section{Conclusions}\label{sec:conclusions}
In conclusion, in this work we have considered the informational limits on image transmission through complex media using a random scattering matrix based formalism. Information-theoretic quantities, namely the mutual information and channel capacity, were considered in preference to more conventional imaging metrics due to the inherent randomisation, and resulting poor image fidelity, caused by scattering in such media. Through Monte Carlo simulations of media described by the COE, CUE and DMPK matrix ensembles, we have numerically studied the full statistical distribution of these metrics and demonstrated the existence of CLTs in the asymptotic limit of large mode numbers. Formal existence conditions for such CLTs were also highlighted. 

Whilst such numerical and ensemble specific results are both interesting and useful, they are nevertheless limited in scope. In this work, we have therefore  established universal upper and lower bounds on the mean mutual information and channel capacity of image transmission through a complex medium. Specifically, the lower bound was found to match that of a fixed transmittance deterministic channel, whereas the upper bound corresponds to a mixture of bimodal channels. For systems with a large number of degrees of freedom, the upper bound on channel capacity was found to be well approximated by that of a bimodal channel with independent identically Bernoulli distributed transmission eigenvalues. Bounds on the variance of the channel capacity were also derived, albeit found to provide only loose bounds for the numerical cases considered since limiting values of the variance are achieved when the channel capacity is Bernoulli distributed. Notably, the limits found here do not require any \emph{a priori} statistical knowledge of the medium other than the mean transmittance and are applicable beyond the more usual diffusive regimes considered in the literature. Given their ensemble independent nature, these bounds hence act as fundamental limits in imaging through scattering media and provide a benchmark to evaluate the many emerging techniques for imaging through complex media.

\section*{Funding}
The Royal Society (UK)

\appendix

\section{Convexity of mutual information and channel capacity with respect to transmission eigenvalues\label{app:convex}}

To prove convexity of the mutual information $I_{\mathcal{N}}$ with respect to the parameters $\eta_j$ we show that its Hessian matrix $\mathbb{H}_{\boldeta}$ is positive semi-definite. We must thus evaluate the derivatives $\partial I_{\mathcal{N}}/\partial \eta_j\partial \eta_k$. Note that we drop the $\mathcal{N}$ notation throughout this section for clarity. Consider then the first order derivative
\begin{align}
\partialdiff{ I_{\mathcal{N}}}{\eta_k} &=\sum_{j=1}^N \partialdiff{}{\eta_k}\left[- p_j \log p_j  + \eta_j p_j \log P_j\right]\\
&= \sum_{j=1}^N \left[\delta_{jk} \,p_j \log P_j + \frac{\eta_j p_j }{P_j} \partialdiff{P_j}{\eta_k} \right]\\
&= p_k \log P_k + \Lambda \sum_{j=1}^N \partialdiff{P_j}{\eta_k} = p_k \log P_k,
\end{align}
where we have used the derivatives $\partial \Lambda_l/\partial p_k = \eta_k$ and $\partial P_j/\partial p_k = \eta_j[\delta_{jk}/\Lambda_l - p_j \eta_k/\Lambda_l^2]$ and the last step follows since $\sum_{j=1}^N \partial{P_j}/{\partial\eta_k} = \partial [\sum_{j=1}^N {P_j}]/\partial{\eta_k}$ where $\sum_{j=1}^N {P_j}=1 $ is a constant. Since $p_j$ and $ P_j $ lie in the range $[0,1]$ it follows that the first derivative is always negative, i.e. the mutual information is a decreasing function with respect to all $\eta_j$. The second order derivative thus takes the form
\begin{align}
\partialdiff{^2 I_{\mathcal{N}}}{\eta_k \partial \eta_l} &= \frac{p_k}{P_k}  \partialdiff{P_k}{\eta_l}=\frac{p_k}{P_k}  \partialdiff{}{\eta_l}\left[\frac{\eta_k p_k}{\sum_{j=1}^N \eta_j p_j }\right]\\
&=  \frac{p_k^2}{P_k}\left[\frac{\delta_{kl}}{\Lambda} - \frac{\eta_k}{\Lambda^2}p_k\right]\\
&= \frac{p_k}{\eta_k}\left[\delta_{kl} - \frac{\eta_k p_l}{\Lambda}\right].
\end{align}
Consider then 
\begin{align}
\mathbf{x}^T \mathbb{H}_{\boldsymbol{\eta}} \mathbf{x} &= \sum_{k=1}^N \sum_{l=1}^N x_kx_l \partialdiff{^2 I_{\mathcal{N}}}{\eta_k \partial \eta_l}\\
&= \frac{1}{\Lambda}\left[\sum_{k=1}^N \eta_k p_k \sum_{l=1}^N \frac{p_l}{\eta_l} x_l^2  - \left(\sum_{k=1}^N x_k p_k\right)^2\right].
\end{align}
From the Cauchy-Schwarz inequality we however note that 
\begin{align}
\sum_{k=1}^N \eta_k p_k \sum_{l=1}^N \frac{p_l}{\eta_l}x_l^2 \geq \left( \sum_{k=1}^N p_k x_k \right)^2
\end{align}
such that $\mathbf{x}^T \mathbb{H}_{\boldsymbol{\eta}} \mathbf{x} \geq 0$, i.e. the Hessian matrix is positive semi-definite. Since the Hessian matrix is positive semi-definite we have that the mutual information $I_{\mathcal{N}}$ is a convex function in $\boldsymbol{\eta}$ for any fixed set of probabilities $\{p_j\}$. At this point we can however use the result that if $f_1(x), f_2(x), \ldots f_N(x)$ are some convex functions in $x$ then their point-wise maximum (i.e. $\mbox{sup}_{\{x\}} f_j$) is also convex in $x$ \cite{Boyd2010}. Accordingly it follows that $C_{\mathcal{N}}$ is a convex function in $\boldsymbol{\eta}$ since it is given by the supremum of $I_{\mathcal{N}}$ with respect to the source probabilities.

\section{Capacity of a bimodal information channel\label{app:bimodal}}
In this section we determine the average channel capacity for scattering media for which the eigenvalues are independent identically distributed Bernoulli random variables. As part of our derivation we will also find the channel capacity for an information channel with $K$ open sub-channels and $N-K$ closed channels. Our derivation here also serves as an illustration as to how to determine the channel capacity in the case when some $\eta_i^{\mathcal{N}}$ are exactly equal to unity and/or zero.

We begin by considering the mutual information of an information channel as given by \eqref{eq:IN} of the main text which takes the form
\begin{align}
I_{\mathcal{N}} &= H(\mathcal{S}) +  \Lambda_{\mathcal{N}}\sum_{j=1}^N P_j^{{\mathcal{N}}}\log P^{{\mathcal{N}}}_j.
\end{align} 
To maximise $I_{\mathcal{N}}$ with respect to the source probabilities $p_j$ subject to the constraint $\sum_{j=1}^N p_j= 1$ we construct the Lagrangian
\begin{align}
L = I_\mathcal{N} + \alpha \left(\sum_{j=1}^N p_j - 1\right)
\end{align}
where $\alpha$ is a Lagrange multiplier.
Evaluating the derivative with respect to $p_k$ yields
\begin{align}
\partialdiff{L}{p_k} &= -(1+\log p_k) + \partialdiff{\Lambda_{\mathcal{N}}}{p_k} \sum_{j=1}^N P_j^{{\mathcal{N}}}\log P^{{\mathcal{N}}}_j \nonumber\\
& \quad + \Lambda_{\mathcal{N}} \sum_{j=1}^N (1+\log P_j^{\mathcal{N}}) \partialdiff{P_j^{\mathcal{N}}}{p_k}\label{eq:dLdpk}
\end{align}
We now assume that $\eta_k^{\mathcal{N}} = 1$ or 0 for all $k$ and that we have ordered the sub-channels such that
\begin{align}
\eta_k^{\mathcal{N}} = \left\{\begin{array}{ll} 
1&\mbox{for~} k \leq m\\
0&\mbox{otherwise}
\end{array} \right. \label{eq:etaorder}
\end{align}
i.e. that there are $m$ closed sub-channels and $K = N-m$ open sub-channels. As in the main text, in this case we define $\mathbf{u} = [\eta_1^{\mathcal{N}},\ldots,\eta^{\mathcal{N}}_N]$.  Accordingly it then follows that
\begin{align}
\partialdiff{\Lambda_{\mathcal{N}}}{p_k} = \left\{\begin{array}{ll} 
1&\mbox{for~} k \leq m \\
0&\mbox{otherwise}
\end{array} \right.
\end{align}
and 
\begin{align}
\partialdiff{P_j^{\mathcal{N}}}{p_k} = \left\{\begin{array}{ll} 
\left[\delta_{jk} - {p_j}/{\Lambda_{\mathcal{N}}}\right]/\Lambda_{\mathcal{N}}&\mbox{for~}  k \leq m\\
0&\mbox{otherwise}
\end{array} \right.
\end{align}
Upon substitution of these derivatives into \eqref{eq:dLdpk} and equating the derivative of the Lagrangian to zero we find
\begin{align}
1-\alpha = \left\{\begin{array}{ll} -\log p_k + \log P_k &\mbox{for~} k\leq m\\
-\log p_k &\mbox{otherwise} \end{array} \right. \label{eq:alphacon}.
\end{align}
Summing \eqref{eq:alphacon} over $k$ and enforcing the constraint $\sum_{k=1}^N p_k=1$ gives
\begin{align}
1-\alpha = -\sum_{k=1}^N \widetilde{p}_k^{\mathcal{N}} \log \widetilde{p}_k^{\mathcal{N}} + \sum_{k=1}^m \widetilde{p}_k^{\mathcal{N}} \log \widetilde{P}_k^{\mathcal{N}} = C_{\mathcal{N}}(\mathbf{u}) \label{eq:alphacon2}
\end{align}
where we have introduced the tilde notation to denote optimal source probabilities which are also dependent on which alphabet ${\mathcal{N}}$ we measure. To determine the optimal probabilities we use \eqref{eq:alphacon} and \eqref{eq:alphacon2} yielding $\widetilde{p}_k^{\mathcal{N}} = \exp[-C_{\mathcal{N}}(\mathbf{u})]$ for $k > m$ and 
\begin{align}
\sum_{k=1}^m \widetilde{p}_k^{\mathcal{N}} = \exp[-C_{\mathcal{N}}(\mathbf{u})]. \label{eq:pksumC}
\end{align}
Although \eqref{eq:pksumC} does not give explicit or unique values for $\widetilde{p}_k^{\mathcal{N}}$ ($k\leq m$), this is of little importance since the $k\leq m$ modes are those which are output in the aggregate mode. Since individual modes can not be distinguished in the aggregate sub-channel, the weightings of the input modes are immaterial. It then follows that
\begin{align}
\sum_{k=1}^N \widetilde{p}_k^{\mathcal{N}} = 1  
&= \sum_{k=1}^m \widetilde{p}_k^{\mathcal{N}} + \sum_{k=m+1}^N \widetilde{p}_k^{\mathcal{N}} \\
&= (1-\delta_{m0}) e^{-C_{\mathcal{N}}(\mathbf{u})} + (N-m )e^{-C_{\mathcal{N}}(\mathbf{u})}
\end{align}
whereby 
\begin{align}
C_{\mathcal{N}}(\mathbf{u}) = \log[N-m + 1-\delta_{m0}]. \label{eq:Cgivenm}
\end{align}
Although we have assumed the specific ordering of $\eta_k^{\mathcal{N}}$ as given by \eqref{eq:etaorder} the derivation is unaffected upon permutation of elements of $\mathbf{u}$. 

We note that thus far in our derivation we have assumed a fixed $\mathbf{u}$ and thus have not allowed for any randomness in our bimodal information channel. For the case that the transmission eigenvalues are independent identically distributed Bernoulli random variables, $m$ is a binomial random variable with corresponding PDF
\begin{align}
p(m) = \sum_{j=0}^N {{N}\choose{j}} \bar{\eta}^j (1-\bar{\eta})^{N-j} \delta(m-j),
\end{align}
where $\delta(\cdots)$ is the Dirac delta function. Note that since the transmission eigenvalues are identically distributed $\bar{\eta}_j = \bar{\eta}$ for all $j$. Determination of the mean capacity of a bimodal channel then follows simply as
\begin{align}
\bar{C}_{\mathcal{N}}^b = \sum_{j=0}^N {{N}\choose{j}} \bar{\eta}^j (1-\bar{\eta})^{N-j} \log[N-j+1-\delta_{j0}].
\end{align}

\section{Central limit theorem for non-linear statistics of $\eta_j$\label{app:CLT}} 
In Ref.~\cite{Politzer1989} Politzer presents a formal proof that the asymptotic probability distribution function of any linear statistic $A = \sum_i \mu(\eta_i)$ of the eigenvalues, here denoted $\eta_i$ ($i=1,\ldots N$), is Gaussian. In his proof Politzer describes how correlations between  eigenvalues can be interpreted as $N$-body forces.  In particular a random ensemble of matrices with $N$-eigenvalue forces can be expressed such that the probability of a set of eigenvalues $\{\eta_j\}$ is proportional to 
\begin{align}
\prod_{i< j} \left|\eta_i - \eta_j\right| \exp \left[\sum_k V(\eta_k)\right] \label{eq:Politzer1}
\end{align}
where $V(\eta_j)$ are effective one body external potentials chosen such that the ensemble has the same eigenvalue density $\rho(\eta)$ and two point correlation function $K(\eta,\eta')$ as the ensemble with the original $N$-body forces. Polizter then proceeds to consider perturbation of the eigenvalue probability distribution by an additional factor of $\exp[\sum_i \mu(\eta_i)]$ where $A = \sum_i \mu(\eta_i)$ such that the eigenvalue density is modified to $\rho(\eta) + \delta \rho(\eta)$. The final step of Politzer's proof is to show that the perturbation in the eigenvalue density $\delta \rho$ is linear in $\mu$, such that the central limit theorem applied. In our case, we can follow analogous steps, however, we now perturb the eigenvalue probability distribution of \eqref{eq:Politzer1} by a nonlinear statistic of the form $A = \sum_i h_i(\boldeta)$, which again perturbs the eigenvalue density to $\rho'(\eta) = \rho(\eta) + \delta \rho(\eta)$. This nonlinear perturbation corresponds to introduction of a perturbing potential with complicated $N$-body forces. Following the arguments of Politzer used to justify the form of \eqref{eq:Politzer1}, it is, however,  possible to replace the perturbed ensemble (when certain smoothness criteria are meet \cite{Kamien1988}) with one with probability distribution of the form 
\begin{align}
\prod_{i< j} \left|\eta_i - \eta_j\right| \exp\left[\sum_k V(\eta_k) + \mu(\eta_k)\right] \label{eq:Politzer2}
\end{align}
whilst maintaining the form of $\rho'(\eta)$ and the perturbed correlation function up to order $1/N$. With this linearised form, the proof of the CLT proceeds identically to that given in Ref.~\cite{Politzer1989}.

\section{Upper informational bounds\label{app:bound}}
In this section we seek to prove that \eqref{eq:CNbarUB} of the main text gives an upper bound no worse than $\log N$ and similarly the analogous expression for the mutual information $I_{\mathcal{N}}$ is no worse than the source entropy $H(\mathcal{S})$. We recall \eqref{eq:CNbarUB} which takes the form
\begin{align}
\bar{C}_{\mathcal{N}}\leq \sum_{\mathbf{u} \in P} C_{\mathcal{N}}\left(\mathbf{u}\right) E_{\boldsymbol{\eta}}\Bigg[\prod_{i \in A_\mathbf{u}  } (1-\eta_i) \prod_{j \in B_\mathbf{u}} \eta_j \Bigg].
\end{align}
We first note that for any $\mathbf{u}$ the inequality $C_{\mathcal{N}}(\mathbf{u}) \leq \log N$ holds, which in turn allows us to factor this term out of the summation such that  
\begin{align}
\bar{C}_{\mathcal{N}}\leq \log N \sum_{\mathbf{u} \in P} E_{\boldsymbol{\eta}}\Bigg[\prod_{i \in A_\mathbf{u}  } (1-\eta_i) \prod_{j \in B_\mathbf{u}} \eta_j \Bigg].
\end{align}
Exchanging the order of summation and expectation we have
\begin{align}
\bar{C}_{\mathcal{N}}\leq \log N \,  E_{\boldsymbol{\eta}} \sum_{\mathbf{u} \in P} \Bigg[\prod_{i \in A_\mathbf{u}  } (1-\eta_i) \prod_{j \in B_\mathbf{u}} \eta_j \Bigg].
\end{align}
Study of the combinatorics of the summation and product terms quickly reveals that 
\begin{align}
\sum_{\mathbf{u} \in P} \Bigg[\prod_{i \in A_\mathbf{u}  } (1-\eta_i) \prod_{j \in B_\mathbf{u}} \eta_j \Bigg] = 1
\end{align}
such that $\bar{C}_{\mathcal{N}} \leq \log N$. The derivation for $I_{\mathcal{N}}$ is formally equivalent except the initial step requires the inequality $I_{\mathcal{N}}(\mathbf{u}) \leq H(\mathcal{S})$.

\section{Maximum channel capacity \label{app:capacity}}
In the main text we derived the ensemble specific upper bound on the mean channel capacity of a scattering medium, as described by \eqref{eq:CNbarUB2}. Specifically we demonstrated that $\bar{C}_{\mathcal{N}} \leq  \mathcal{C}$ where
\begin{align}
\mathcal{C} =\sum_{K=0}^N \widetilde{w}_K \log[K + 1 - \delta_{KN}].
\end{align}
In this section we show that among all possible ensembles with a fixed mean total  transmittance the universal upper bound is given by $\sup_{\{\widetilde{w}_K \}}  \mathcal{C}= \bar{C}_{\mathcal{N}}^{\txtpow{max}}$, where $\bar{C}_{\mathcal{N}}^{\txtpow{max}}$ is given by \eqref{eq:C_GUB} of the main text. Specifically we note that the maximisation of $\mathcal{C}$ is subject to the constraints
\begin{align}
\sum_{K=0}^N \widetilde{w}_K &= 1\label{eq:derivcon2}
\end{align}\begin{align}
\sum_{K=0}^N (N-K)\widetilde{w}_K &= \bar{g}_{\mathcal{N}}\label{eq:derivcon3}.
\end{align}
where
\begin{align}
0\leq \widetilde{w}_K \leq {{N}\choose{K}} \quad \mbox{for all~}K.\label{eq:derivcon1}
\end{align}
Although the constraint given in \eqref{eq:derivcon1} was not explicitly given in the main text it follows by observing that $0\leq w(\mathbf{u})\leq 1$, $\widetilde{w}_K  = \sum_{\mathbf{u}\in U_K} w(\mathbf{u})$ and that the set $U_K$ has $\#U_K = {{N}\choose{K}}$ distinct elements. \eqref{eq:derivcon2} and \eqref{eq:derivcon3} can be used to eliminate two of the $\widetilde{w}_K$ from the complete set of $N$. Specifically we considering expressing $\widetilde{w}_i$ and $\widetilde{w}_{i-1}$ in terms of the remaining $\widetilde{w}_K$, which from \eqref{eq:derivcon2} and \eqref{eq:derivcon3} gives
\begin{align}
\widetilde{w}_i &= s_2-\bar{g}_{\mathcal{N}}  + (N-i+1)(1-s_1)\\
\widetilde{w}_{i-1} &=  \bar{g}_{\mathcal{N}} - s_2 + (i-N)(1-s_1)
\end{align}
where 
\begin{align}
s_1 &= \sum_{\substack{K=1\\ K\neq i,i-1}}^N \widetilde{w}_K \\
s_2 &= \sum_{\substack{K=1\\ K\neq i,i-1}}^N (N-K)\widetilde{w}_K.
\end{align}
We now consider the explicit difference
\begin{align}
\Delta = \bar{C}_{\mathcal{N}}^{\txtpow{max}} - \mathcal{C}.
\end{align}
Setting $i = N-k$, where $k = \mbox{floor}[\bar{g}_{\mathcal{N}}]$, gives
\begin{align}
\Delta =-\sum_{\substack{K=1\\ K\neq i,i-1}}^N \widetilde{w}_K &\Big[\log(K+1-\delta_{KN})  + (K-i)\log(i)\nonumber\\
&\quad+ (i-K-1)\log(i+1-\delta_{iN}) \Big].\label{eq:Delta}
\end{align}
From \eqref{eq:Delta} it is first observed that when $\widetilde{w}_K = 0$ for $\{K;0\leq K\leq N, K\neq N-k,N-k-1\}$ the difference between $\mathcal{C}$ and and $\bar{C}_{\mathcal{N}}^{\txtpow{max}}$ is identically zero. Noting further that the bracketed factor in \eqref{eq:Delta} is negative for all $0\leq K\leq N$ and $0\leq k \leq N-1$ (as can be easily seen by visual inspection of the function), it follows that for any $\widetilde{w}_K \geq 0$ ($\{K;0\leq K\leq N, K\neq N-k,N-k-1\}$) the difference $\Delta$ is positive, i.e. $\mathcal{C}\leq \bar{C}^{\txtpow{max}}_{\mathcal{N}}$. Positivity of $\widetilde{w}_K$ therefore ensures that $\bar{C}^{\txtpow{max}}_{\mathcal{N}}$ represents the universal upper bound on the mean channel capacity.

\end{document}